\documentclass[11pt,twoside]{article}

\usepackage{epsf}

\newcommand{\ul}[1]{\underline{#1}}

\newcommand{\rmj}{{\rm j}}

\newcommand{\rmd}{{\rm d}}

\newcommand{\bea}{\begin{eqnarray}}

\newcommand{\eea}{\end{eqnarray}}

\begin{document}


\title{Angular Spectral Plane-Wave Expansion of\\ Nonstationary Random Fields in\\ Stochastic Mode-Stirred Reverberation Processes
}

\author{{Luk R. Arnaut}\\
\it \small Time, Quantum and Electromagnetics Division,\\
\it \small National Physical Laboratory,\\
\it \small Teddington TW11 0LW,
United Kingdom\\
\rm \small and\\
\it \small Department of Electrical and Electronic Engineering,\\
\it \small Imperial College of Science, Technology and Medicine,\\
\it \small South Kensington Campus, London SW7 2AZ, United Kingdom
}%

\date{\today}

\maketitle

\begin{abstract}
We derive an integral expression for the plane-wave expansion of the time-varying (nonstationary) random field inside a mode-stirred reverberation chamber. It is shown that this expansion is a so-called oscillatory process, whose kernel can be expressed explicitly in closed form. The effect of nonstationarity is a modulation of the spectral field on a time scale that is a function of the cavity relaxation time. It is also shown how the contribution by a nonzero initial value of the field can be incorporated into the expansion. The results are extended to a special class of second-order processes, relevant to the perception of a mode-stirred reverberation field by a device under test with a first-order (relaxation-type) frequency response. 
\end{abstract}

\section{Introduction \label{sec:intro}}
In recent years, complex scenarios and applications in electromagnetism have sparked a renewed interest in the theory and characterization of random electromagnetic (EM) fields in static and dynamic environments. 
Examples include characterization of the impedance and radiation efficiency of electrically small antennas \cite{kild1}, 
multi-terminal wireless communications systems in multi-path high-mobility propagation environments \cite{migl1}, dynamic atmospheric propagation effects \cite{chri1} on the group delay for satellite-based navigation systems, radio-frequency and optical speckle and scintillation associated with interaction of EM waves with rough surfaces or random media \cite{yura1}, parametrically varying microwave billiards and cavities \cite{diet1}--\cite{rosn2}, electromagnetic compatibility (EMC) including testing of emissions or immunity of electronic equipment to irradiating EM fields \cite{arna_timedomain}, etc.
Continuously evolving multi-scattering electromagnetic environments (EMEs) are ubiquitous but difficult to characterize and analyze accurately, even using full-wave numerical simulation methods, because of their large size relative to the wavelength and consequent extreme sensivity to small perturbations or configurational uncertainties, and because of complicated transient effects occurring in the propagating signals that result from dynamically changing boundary conditions, especially in resonant multi-scatter environments \cite{rosn2}--\cite{arna_timedomain}.
Because of the spatial and/or temporal fluctuations of the wave properties in such an EME, their characteristics can often be considered as quasi-random and can be efficiently modelled using stochastic methods. From a physics point of view, the task is then to characterize the statistical parameters and distributions of the random field in terms of deterministic physical quantities of the problem of interest.

A canonical EME for generating statistically random vector fields is the mode-tuned or mode-stirred reverberation chamber (MT/MSRC) \cite{arnaPRL}, \cite{lamb1}. It consists of an electrically large cavity furnished with a mechanical stirring mechanism to alter the boundary and/or excitation conditions (e.g., a large reflective paddle wheel or moving walls) or a varying (hopping or swept) excitation frequency or noise source for mixing the modal fields. Such cavities are important devices for studing the temporal, spatial, structural, and polarization properties of local instantaneous scalar and vector wave fields, as well as characteristics of their spatio-temporal coherence (correlation) properties. 

For a static or quasi-static MT/MSRC, the assumption of wide-sense quasi-stationarity of the field is an excellent and often-made approximation.
However, when the rate of change of the EME (expressed by the correlation time in space or time) is relatively fast compared to the characteristic time or length scale of the field in the corresponding static EME (e.g., frequency, relaxation time), this assumption breaks down \cite{arnaTEMCv47n4}. In this case, more sophisticated field models and methods of analysis are necessary. The purpose of the present paper is to develop a framework for the theoretical characterization of such nonstationary fields.

The issue bears resemblance to the filtered observation of dynamic scenes through natural vision. The limited image resolution time of the human naked eye ($\sim$ 30 ms) and processing time of vision by the brain ($\sim$ 150 ms) limits the amount and capacity of visual information that can be faithfully captured in real time. In particular, ultra-short-term transient natural phenomena may therefore be perceived as distorted (blurred) or even completely masked. This is a manifestation of (nonlinear) nonstationary filtering of the actual scene. Using high-speed cameras, the image frames and dynamics of the hidden reality can be reconstructed. Parenthetically, compound eyes allow for detecting transients more rapidly, through high temporal resolution of flicker across the ommatidia, but at the expense of poorer spatial resolution compared to single-chambered eyes \cite{wils1}, \cite{voel1}. 

\section{Methods for expansion of vector EM fields in dynamic complex confined environments}
For the representation of fields inside a MSRC or in a complex EME exhibiting multiple scattering or multipath fading, two methods have found application. Under idealized conditions, both methods should lead to equivalent results, but each has its particular advantages and disadvantages.
{\it Modal expansions} (MEs) consider the EME as a multi-mode system characterized by a set of vector eigenmodes $\ul{\psi}_{mnp}$ at discrete resonance frequencies (eigenfrequencies) $\omega_{mnp}$ at time $t$. These eigenmodes form a set of basis functions for expanding the local stationary interior field, with expansion coefficients $a_{mnp}$. In a dynamic EME, the eigenmodes and expansion coefficients both depend on time \cite{tai1}:
\bea
\ul{E}(\ul{r},t) = \sum_{m,n,p} a_{mnp}(t) \ul{\psi}_{mnp}(t).
\eea
MEs are exact, in the sense of inherently incorporating the EM boundary conditions and cavity geometry into the description. Since modes are nonlocal, the description inherently includes the spatial inhomogeneity of the field caused by the vicinity of boundaries and objects. Since a ME does not presume quasi-randomness of the field, it can be used in principle at any frequency, although the increasing modal count and modal overlap (modal coupling) in high-Q chambers for increasing frequencies may make their implementation prohibitive at very high frequencies. 
A greater burden is the fact that realistic chambers often exhibit complex-shaped time-varying boundaries (diffractors, e.g., corrugations and mode stirrers), which hampers the practical implementation of MEs that require the calculation and/or functional form of the eigenmodes. 
Furthermore, a correct description requires the cavity to be described as a partially opened system \cite{ditt1}, due to the transmitter and receiver introducing apertures to the unbounded exterior region of the cavity. This requires mixed (Robin) boundary conditions and complicates the analysis and eigenmode characterization.

In a second method, {\it angular spectral plane-wave expansions} (ASPWEs) have been used for deterministic \cite{whit1}, \cite{clem1} and random \cite{schroeder}, \cite{berr1}, \cite{dunn1} fields.
In ASPWEs, the local field with a suppressed $\exp(\rmj \omega t)$ time dependence for its carrier is expanded as a discrete sum (for bounded closed finite-sized cavities) or integral (for unbounded opened regions supporting quasi-random fields) of an angular spectrum of plane waves \cite{whit1}, \cite{hill1}, \cite{arnaPRE1}, \cite{arnaRS}:
\bea
\ul{E}(\ul{r},t) = \frac{1}{\Omega} \int\int_{\Omega} \ul{\cal {E}}(\Omega,t) \exp \left [ -\rmj \ul{k}(t) \cdot \ul{r} \right ] {\rm d}\Omega
\eea
where $\Omega = 2\pi$ sr or $4\pi$ sr is the solid angle associated with a half-space or entire space for the incident field, respectively.
Unlike MEs, ASPWEs provide characterizations of the local\footnote{If the field is statistically homogeneous (spatially uniform), then the characterization applies of course throughout the volume.} field, thus avoiding the need for determining eigenmodes explicitly. They lend themselves easily to statistical characterization of this local field and, when extented to spatial correlation functions, to nonlocality within a finite region.
With the aid of proposed axioms for the average and covariance of the amplitude spectrum \cite{chri1}, \cite{vanm1}, \cite{yagl1} and spectral density \cite{hill1} of the wave components, explicit calculation of average and covariance of the expanded random field is possible.
By {\it a posteriori} imposing EM boundary conditions, the ASPWE even enables a complete derivation of probability density functions based on the statistical moments \cite{arnaRS}, \cite{arnaPEC1}.

In both MEs and ASPWEs, integral representations are used as approximations to discrete sums in cavities that are sufficiently large relative to the wavelength.
Strictly, both methods are applicable only for static or quasi-statically perturbed cavities, for which the concept of modes and their decomposition into standing plane waves are properly defined. When boundaries are moved, complications arise relating to nonstationary transient fields and dynamics of eigenmodes, as well as the very concept and definition of eigenmodes in such circumstances.
Note that, in order for the effect of this motion to be significant, this does not necessitate relativisitic speeds, unlike in unbounded EMEs:
 the relevant time scales are the {\em ratios\/} of the correlation time of the mode-stirring process to the decay constant of the cavity (at a given frequency), to the response time of the antenna or to the characteristic time (e.g., cycle period) of the device under test (DUT), and of course to the period of the source wave itself \cite{arnapaddletransients1}. For example, it is well known that even small Doppler shifts (cf. Sec. \ref{sec:specexpan}), frequency changes of the order of a few hundred Hz at $1$ GHz may cause degradation of performance of wireless communication systems, because they give rise to considerable jitter and distortion in impulse radio and other wideband signals.

In this paper, we develop an extension of the ASPWE for steady-state fields to nonstationary fields. The analysis is for temporal nonstationarity of scalar fields; corresponding results for spatial nonstationarity and vector fields follow mutatis mutandis.

\section{Reverberation process with stirring slip}
\subsection{Spectral expansion\label{sec:specexpan}}
A model of a slipping random field based on a random-walk model was presented in \cite{arnaJPA}, to which we refer for details. Here we briefly review and summarize the main results of this model. 

\begin{figure}[htb] \begin{center} \begin{tabular}{l}
\ \epsfxsize=16cm 
\epsfbox{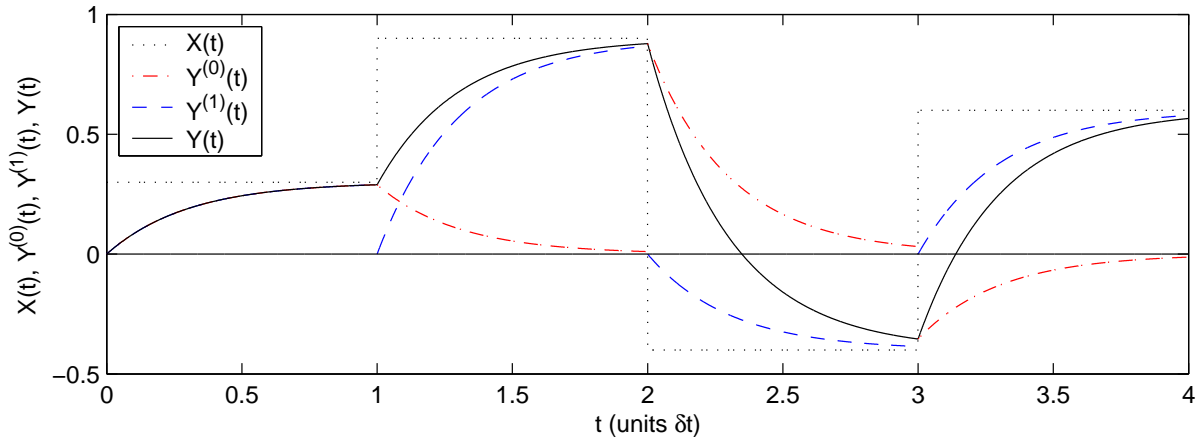}\ \\
\end{tabular}
\end{center}
{
\caption{\label{fig:nonstationary_diagram} \small
(color online)
Discretized sample-and-hold input process $X(t)$ and resulting weighted mean field $Y(t)=Y^{(0)}(t)+Y^{(1)}(t)$ for a first-order process, with time-independent average relaxation time $\tau=0.3\thinspace\delta$t (nonlinear weighting). The actual continuous-time stirring process is obtained as the limit $\delta t \rightarrow 0$.}}
\end{figure}

We consider the evolution of a nonstationary detected statistical mean field $Y(t)$, resulting from a stationary source field $X(t)$ (Fig. \ref{fig:nonstationary_diagram}). Here, $X(t)$ is taken to be a periodic field with deterministic waveform, as emitted inside a static cavity. (By extension, a wide-sense stationary noise could be treated.) In the application to EMC in Sec. \ref{sec:applic}, $X(t)$ and $Y(t)$ are modulated complex RF fields. The field $Y(t)$ is produced by random step changes $\delta X(t)$ induced by a change in boundary conditions and weighted in the mean by a relaxation time, i.e., it is governed by the first-order Langevin--It\^{o} stochastic differential equation (SDE): 
\bea
\frac{\rmd Y(t)}{\rmd t} + \frac{1}{\tau} Y(t) = \frac{1}{\tau} X(t).
\eea
This equation of motion for the random field spans a variety of cases of field dynamics across a time interval $\Delta t$, ranging from instantaneous response ($\Delta t/\tau \rightarrow +\infty$) to Brownian motion ($\Delta t/\tau \rightarrow 0$).
Within each infinitesimal time increment $\delta t$, $Y(t)$ evolves from its initial value at $t_0$ in the direction of its asymptotic steady-state value that would be reached for $\Delta t \rightarrow +\infty$. 
In Fig. \ref{fig:nonstationary_diagram}, $Y^{(0)}(t_i)$ for $t_i \in [t_0,t_1]$ represents the mean outgoing (fading) field for $t_i$, i.e., represented by the set of plane-wave components associated with the ``previous'' state of the system and evaluated at $t=t_{i-}$. This ``previous'' field decays at a rate governed by the decay time $\tau$ for each plane-wave component. By contrast, $Y^{(1)}(t_i)$ at any $t_i \in [t_0,t_1]$ represents the mean incoming (emerging) field at $t_i$, i.e., associated with the set of plane-wave components for the ``next'' state, as evaluated at $t=t_{i+}$. The fact that the outgoing field requires a nonzero time to vanish gives rise to a ``slipping'' field, i.e., a process with a gradually fading memory, such that $Y(t)$ is a nonuniformly weighted mixture of both contributions, viz., $Y(t_0+\Delta t) = Y^{(1)}(t_0+\Delta t) + c(\Delta t) Y^{(0)}(t_0)$ with $0\leq |c(\Delta t)|\leq 1$ (cf. \cite{arnaJPA} for a detailed classification and discussion).

The fact that this evolution can be characterized, in the mean, by a single time constant $\tau$ governing the rate of change of the instantaneous mean envelope of the transient field (mean effective relaxation time $\tau$) has been amply supported by numerical and practical experiments \cite{arnaPRL}, \cite{rich1}, \cite{kwon1}. In actuality, each component of the field, whether angular-spectral or modal, has its own particular relaxation time constant associated with it. However, their values are relatively close within a narrow band of frequencies, in the sense that
\bea
\frac{| \tau_{mnp} - \tau_{m^\prime n^\prime p^\prime} | }{ \tau_{mnp} + \tau_{m^\prime n^\prime p^\prime} } \ll 1
\eea
for any pair of modes (or standing plane waves with oppositely directed wave vectors) $(m,n,p)$ and $(m^\prime, n^\prime, p^\prime)$ located within this frequency band.
Furthermore, strong modal coupling, both in time and frequency, causes equalization of their different values $\tau_{mnp}(f_{mnp},t_i)$. For these reasons, a one-parameter description is often sufficiently adequate, at least in a mean-square sense, i.e., when the field arises as some weighted average, as observed in a coupled multi-modal incoherent system. 
In any case, extensions of the model are possible. For example, a second-order model governing the response to step transitions would incorporate not only the average rate of $Y(t)$ approaching its steady-state value, but also the average level of overshoot of $Y(t)$ above steady-state during such transitions.

In \cite[Eq. (9)]{arnaJPA}, the resultant $Y(t)$ in a discretized interval $[t_i, t_{i+1}]$ 
was expressed iteratively, as a function of $X(t_i)$. For the present purpose, it is  more beneficial
to use the following equivalent closed-form expression for $Y(t)$ 
\bea
&~& \hspace{-0.8cm}
Y(t_m \leq t \leq t_{m+1}) \nonumber\\
&=&
Y(t_0) \exp \left ( - \frac{t - t_0}{\tau} \right ) 
+ \exp \left ( - \frac{t - t_m}{\tau} \right ) 
\nonumber\\
&~&
\times
\left [
1 - \exp \left ( - \frac{\delta t}{\tau} \right )
\right ]
\sum^{m-1}_{i=0} \exp \left ( - \frac{t_{m-1} - t_i}{\tau} \right ) X(t_i)
\nonumber\\
&~&
+
\left [ 1 - \exp \left ( - \frac{t - t_{m}}{\tau} \right )
\right ] X(t_m)
\label{eq:Ynoniter}
\eea
for $m\geq 1$, 
where
$
t_m \stackrel{\Delta}{=} t_0 + m \delta t
$.
For $m=0$, the same expression holds, but without the second term.
Neither $Y(t)$ nor its increments $\delta Y_m(t) \equiv Y(t_{m+1})-Y(t_m)$ are in general stationary, because both depend on $t_m$.
Since we shall further be taking the limit $\Delta t / \tau \rightarrow 0$ and because we shall be mainly interested in late times of observation such that $(t-t_0)/\delta t \gg 1$, we restrict further evaluation to the discrete time instances $t=t_{m+1}$, when Eq. (\ref{eq:Ynoniter}) simplifies to
\bea
Y(t_{m+1}) 
&=&
Y(t_0) \exp \left ( - \frac{(m+1) \delta t}{\tau} \right ) + \exp \left ( - \frac{m \thinspace \delta t}{\tau} \right ) 
\nonumber\\
&~&
\times
\left [
1 - \exp \left ( - \frac{\delta t}{\tau} \right )
\right ]
\sum^{m}_{i=0} \exp \left ( + \frac{i \thinspace \delta t}{\tau} \right ) X(t_i)
\label{eq:Ynoniter2}
\eea
Upon taking the limit $\delta t / \tau \rightarrow 0$ in Eq. (\ref{eq:Ynoniter2}), the discrete sum reduces to an integral from $t_0$ to $t_m$, and we obtain
\bea
Y(t_m) &\rightarrow& Y(t_0) \exp \left ( - \frac{t_{m+1} - t_0}{\tau} \right ) 
+ \lim_{\delta t\rightarrow 0} \frac{ 1 - \exp \left ( - \frac{\delta t}{\tau} \right )}{\delta t} \nonumber\\
&~& \times \int^{t_{m}}_{t_0} \exp \left ( - \frac{t_{m}-s}{\tau} \right ) X(s) \rmd s
\label{eq:six}
\eea

At this point, we introduce the Fourier--Stieltjes spectral expansion of the stationary source field $X(t)$ \cite{yagl0}, given by
\bea
X(s) = \int^{+\infty}_{-\infty} \exp \left (\rmj \omega s\right ) \rmd Z_X(\omega)
\label{eq:seven}
\eea
in which $Z_X(\omega)$ for an ideal mode-tuned field is assigned the following properties \cite{vanm1}, \cite{yagl0}:
\bea
\langle \rmd Z_X(\omega) \rangle &=& 0,\\
\langle \rmd Z_X(\omega_1) \rmd Z^*_X(\omega_2) \rangle &=& \delta (\omega_1-\omega_2) \rmd F_X(\omega_{1,2}),
\label{eq:orthog_properties}
\eea
i.e., $Z_X(\omega)$ is orthogonal.
[Note that $Z_X(\omega)$ represents the accumulated (integrated) spectrum of $X$, not to be confused with its differential, i.e., the Fourier spectrum.]
With the substitution $u\stackrel{\Delta}{=}(t_m-s)/\tau$ and letting $t_{m+1}\rightarrow t_m \stackrel{\Delta}{=} t$ in Eq. (\ref{eq:six}), we arrive after some manipulation at
\bea
Y(t) 
&=&
Y(t_0) \exp \left ( - \frac{t-t_0}{\tau} \right ) 
+
\int^{+\infty}_{-\infty} \exp \left ( \rmj \omega t \right )
\nonumber\\
&~& \times
\frac{1-\exp \left [ - \left ( 1 + \rmj \omega \tau \right ) \frac{t-t_0}{\tau} \right ]}{1 + \rmj \omega \tau} \rmd Z_X(\omega). \label{eq:evol1}
\eea
Turning attention to the first term in Eq. (\ref{eq:evol1}), 
and using general properties of the Fourier transform and the Dirac delta distribution \cite[Eqs. (2.24) and (2.40)]{cham1} for $X(t_0) = X(t) \delta (t-t_0)$, this term can be assigned a spectral expansion, viz.,
\bea
&~& \hspace{-0.8cm}
X(t_0) \exp \left ( - \frac{t-t_0}{\tau} \right ) \nonumber\\
&=& \int^{+\infty}_{-\infty} X(t_0) \frac{\tau \exp \left [ \rmj \omega (t-t_0) \right ] }{1 + \rmj \omega \tau} {\rm d}\omega
\\
&=& \int^{+\infty}_{-\infty} \left [ \int^{+\infty}_{-\infty} X(\omega^\prime) \exp \left [ \rmj \omega^\prime (t-t_0)\right ]  {\rm d}\omega^\prime \right ]\nonumber\\
&~&\times \frac{\tau \exp \left [ \rmj \omega (t-t_0) \right ]}{1 + \rmj \omega \tau} {\rm d}\omega .
\eea
Note that $X(t_0)\equiv Y(t_0)$ because nonstationarity has not yet manifested itself at the start time $t_0$, whence the output process can then be completely identified with the input process at this instance. With an interchange of the order of integration, this yields the general result
\bea
Y(t_0) \exp \left ( - \frac{t-t_0}{\tau} \right ) 
&=& \int^{+\infty}_{-\infty} \theta(\omega;t) \exp (\rmj \omega t) {\rm d}Z_X(\omega)
\label{eq:Yintegral_generalsol}
\eea
where
\bea
\theta(\omega;t) \stackrel{\Delta}{=} 
\tau \exp ( -\rmj \omega t_0 ) 
\int^{+\infty}_{-\infty} \frac{\exp \left [ \rmj \omega^\prime (t- t_0 ) \right ] }
                              {1 + \rmj \omega^\prime \tau}  {\rm d}\omega^\prime. \label{eq:theta}
\eea
In the particular case where we impose the initial condition $Y(t_0)=0$, for simplicity, only
the second term in Eq. (\ref{eq:evol1}) remains and can be represented as an {\it oscillatory process} \cite{prie1}, i.e.,
\bea
Y(t) = \int^{+\infty}_{-\infty} \phi(t;\omega,\tau) \exp \left ( \rmj \omega t\right ) {\rm d}Z_X(\omega),
\label{eq:Yintegral_zerostart}
\eea
whose kernel
\bea
\phi(t;\omega,\tau) \stackrel{\Delta}{=} \frac{1-\exp \left [ - \left ( 1 + \rmj \omega \tau \right ) \frac{t-t_0}{\tau} \right ]}{1 + \rmj \omega \tau}
\label{eq:defphi}
\eea
is a complex amplitude modulation function expressed in explicit, i.e., closed form.
It is seen from Eq. (\ref{eq:Yintegral_zerostart}) that the effect of nonstationarity on the plane-wave expansion is a modulation of the spectral field on a time scale that is a function of the cavity relaxation time.

\begin{figure}[htb] \begin{center} \begin{tabular}{c}
\ \epsfxsize=16cm 
\epsfbox{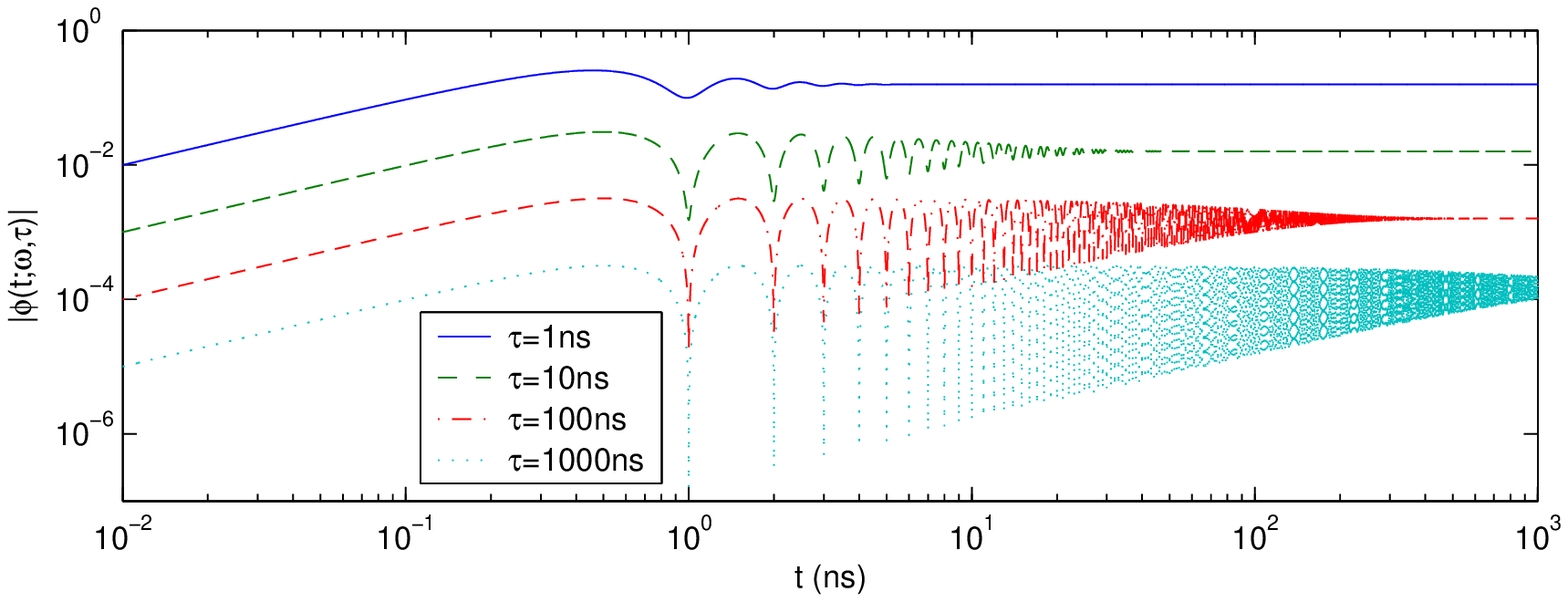}\ \\
(a)\\
\ \epsfxsize=16cm 
\epsfbox{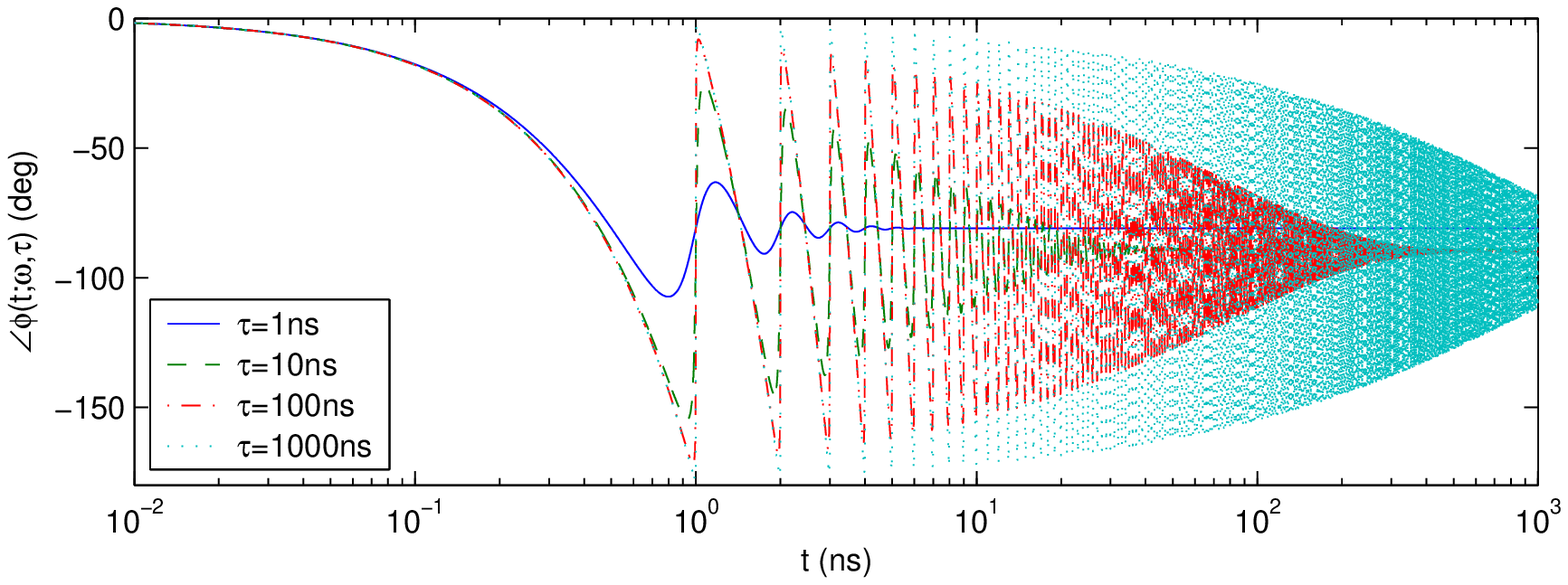}\ \\
(b)
\end{tabular}
\end{center}
{
\caption{\label{fig:nonstationary_kernel} \small
(color online) Modulating function $\phi(t;\omega,\tau)$ of the kernel in Eq. (\ref{eq:Yintegral_zerostart}) for the spectral plane-wave representation of nonstationary mode-stirred fields, for selected values of $\tau$ at $\omega/(2\pi) = 1$ GHz: (a) magnitude; (b) phase.}}
\end{figure}

Fig. \ref{fig:nonstationary_kernel} shows the time evolution of Eq. (\ref{eq:defphi}) at selected values of $\tau=Q/\omega$ for an overmoded cavity, for narrowband operation at an excitation frequency $\omega = 2\pi\times 10^9$ rad/s. It is seen that larger values of $\tau$ (and, hence, $Q$) result in longer transition times before $\phi$ reaches a regime, as is intuitively clear. The strongest variability of $\phi(t)$ occurs when $t \sim 2\pi/\omega$, as expected.

As is well known, oscillatory processes provide an evolutionary spectral representation for a special class of nonstationary processes \cite[Eq. (3.9)]{prie1}, \cite{gran1}. They apply when nonstationarity is weak (i.e., sufficiently slowly evolving), so that the evolution can be considered as a modulation of the original process $X(t)$.  

For $\omega\tau \rightarrow 0$ and $t/\tau \rightarrow +\infty$, we retrieve from Eq. (\ref{eq:evol1}) the classical spectral expansion for temporally stationary (but spatially homogeneous as well as inhomogeneous) fields, viz.,
\bea
Y(t) =
\int^{+\infty}_{-\infty} \exp \left ( {\rm j} \omega t \right ) \rmd Z_X(\omega) \equiv X(t). \label{eq:stat1}
\eea
On the other hand, for $\tau \gg 1/\omega$ we obtain a ${\rm sinc}(\omega t)$-type modulation, in view of the symmetric integration limits.

\subsection{Statistics}
The ensemble mean value of $Y(t)$ is
\bea
\langle Y(t) \rangle
=
Y(t_0) \exp \left ( - \frac{t-t_0}{\tau} \right )
\label{eq:ensemblemean}
\eea
because $\langle {\rm d}Z_X(\omega) \rangle = 0$. 
The covariance function follows from Eq. (\ref{eq:evol1}) as
\bea
&~&\hspace{-0.8cm}
\langle Y(t_1) Y^*(t_2) \rangle\nonumber\\
&=&
|Y(t_0)|^2 \exp \left ( - \frac{t_1+t_2-2t_0}{\tau} \right ) \nonumber\\
&~& + \int^{+\infty}_{-\infty}
\frac{\exp \left [ \rmj \omega (t_1-t_2) \right ]}{1 + \omega^2 \tau^2} \nonumber\\
&~& \times 
\left \{ 1 + \exp \left ( - \frac{t_1 +t_2 - 2 t_0}{\tau} \right ) \exp \left [ - \rmj \omega (t_1-t_2) \right ]
\right. \nonumber\\
&~&~~~\left.
- \exp \left ( - \frac{ t_1 - t_0 }{\tau} \right ) \exp \left [ - \rmj \omega \left ( t_1 - t_0 \right ) \right ] 
\right. \nonumber\\
&~&~~~\left.
- \exp \left ( - \frac{ t_2 - t_0 }{\tau} \right ) \exp \left [ \rmj \omega \left ( t_2 - t_0 \right ) \right ] 
\right \}
\rmd F_X(\omega).
\eea
This general result is important in the investigation of the effect of nonstationarity on the transformation of correlation characteristics and effective number of degrees of freedom.
For $t_1=t_2\stackrel{\Delta}{=}t$, the variance follows as
\bea
\sigma^2_{Y}(t) 
&=& 
\langle |Y(t)|^2 \rangle - | \langle Y(t) \rangle |^2
\nonumber\\
&=&
\int^{+\infty}_{-\infty}
\left [ {1 + \omega^2 \tau^2} \right ]^{-1} \left \{ 1  
- 2 \exp \left ( - \frac{t-t_0}{\tau} \right ) \right. \nonumber\\
&~&
\left.
\times \cos \left [ \omega (t-t_0) \right ] 
+ \exp \left [ - \frac{ 2 (t-t_0)}{\tau} \right ] \right \}
\rmd F_X(\omega),~~~
\label{eq:nineteen}
\eea
which describes the evolution of fluctuation levels under nonstationarity.
We now consider two important special cases for the input process.

\subsubsection{Ideal white noise}
If $X(t)$ is ideal white noise, i.e., $\rmd F_X(\omega) = f_X(\omega) \rmd\omega = \sigma^2_X {\rm d}\omega$, then
from Eqs. (\ref{eq:ensemblemean}), (\ref{eq:nineteen}), and [\cite{grad1}, Eq. (3.723.2)],
\bea
\sigma^2_Y(t) = \frac{\pi \sigma^2_X}{\tau} \left \{ 1 - \exp \left [ - \frac{2(t-t_0)}{\tau} \right ] \right \}.
\label{eq:WGN_var}
\eea
In the limits $t/\tau \rightarrow 0$ and $t/\tau \rightarrow +\infty$, we retrieve $\sigma^2_Y \rightarrow 2 \pi \sigma^2_X (t-t_0) / \tau^2$ and $\sigma^2_Y \rightarrow \pi \sigma^2_X/\tau$, respectively, as expected.

\subsubsection{Ornstein-Uhlenbeck process}
\begin{figure}[htb] \begin{center} \begin{tabular}{l}
\ \epsfxsize=6cm 
\epsfbox{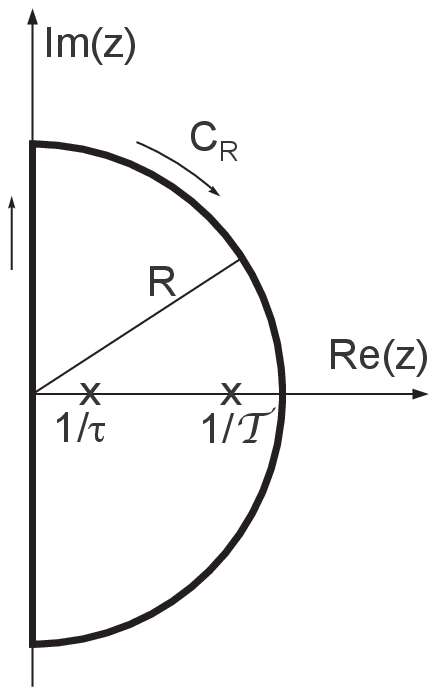}\ \\
\end{tabular}
\end{center}
{
\caption{\label{fig:contourint} \small
Contour of integration for Eq. (\ref{eq:contourint_kernel}) in the complex $z$-plane ($\tau > {\cal T}$).}}
\end{figure}

If $X(t)$ is a first-order process that is wide-sense stationary and exponentially correlated with finite correlation time ${\cal T}$, i.e.,
\bea
\rho_X(t-t_0) = \exp \left ( -\frac{t-t_0}{\cal T} \right )
\eea
so that 
$
\rmd F_X(\omega) = f_X(\omega) \rmd\omega = {2\sigma^2_0{\cal T}}/[{\pi \left ( 1+ \omega^2 {\cal T}^2 \right )}] \rmd\omega 
$,
then Eq. (\ref{eq:nineteen}) reduces to
\bea
\sigma^2_Y(t) &=& \frac{2 \sigma^2_0 {\cal T}}{\pi}
\left \{ 1 + \exp \left [ - \frac{2(t-t_0)}{\tau} \right ] \right \} \nonumber\\
&~& \times \int^{+\infty}_{-\infty} \frac{\rmd \omega}{(1+\omega^2\tau^2)(1+\omega^2{\cal T}^2)} \nonumber\\
&~&
- \frac{4 \sigma^2_0 {\cal T} }{\pi} \exp \left [ - \frac{(t-t_0)}{\tau} \right ] \nonumber\\
&~& \times \int^{+\infty}_{-\infty} \frac{\cos \left [ \omega (t-t_0) \right ] \rmd \omega}{(1+\omega^2\tau^2)(1+\omega^2{\cal T}^2)}.
\label{eq:OU_moment2}
\eea
The first integral in Eq. (\ref{eq:OU_moment2}) is easily calculated with the aid of [\cite{grad1}, Eqs. (2.124.1) and (2.161.1)], yielding
$
{\pi}/({\tau +{\cal T}})
$,
i.e., providing a positive contribution
irrespective of whether $\tau \leq {\cal T}$ or $\tau \geq {\cal T}$.
The second integral in Eq. (\ref{eq:OU_moment2}) is obtained by contour integration of 
\bea
\phi(z) = \frac{\exp \left [ - z (t-t_0) \right ]}{(1-z^2\tau^2)(1-z^2{\cal T}^2)}
\label{eq:contourint_kernel}
\eea
across the left half of the complex $z$-plane (Fig. \ref{fig:contourint}). The residues in $z_1 = 1/\tau$ and $z_2 = 1/{\cal T}$ are
\bea
R_1 = \frac{\tau\exp \left ( - \frac{t-t_0}{\tau} \right )}{2\left ( {\cal T}^2 - \tau^2 \right )}
{~~\rm and~~}
R_2 = \frac{{\cal T}\exp \left ( - \frac{t-t_0}{{\cal T}} \right )}{2\left ( \tau^2 - {\cal T}^2\right )},
\eea
respectively, whence
\bea
\sigma^2_Y(t) = \frac{2\sigma^2_0{\cal T}}{\tau +{\cal T}} \left \{ 1 - \exp \left [ - \frac{2(t-t_0)}{\tau} \right ] \right \}. \label{eq:OU_moment2_bis}
\eea
Again, this result holds irrespective of whether $\tau \leq {\cal T}$ or $\tau \geq {\cal T}$. 
It is verified that for ${\cal T}/\tau \ll 1$, Eq. (\ref{eq:OU_moment2_bis}) reduces to Eq. (\ref{eq:WGN_var}) with $\sigma^2_X = 2 \sigma^2_0 {\cal T}/\pi$.
For $(t-t_0)/\tau \ll 1$, we retrieve the linear dependence as in the case of ideal white noise under the same condition. Furthermore, for ${\cal T}/\tau \rightarrow 0$, it is verified that Eq. (\ref{eq:OU_moment2_bis}) approaches Eq. (\ref{eq:WGN_var}) with $2\sigma_0 {\cal T} = \pi \sigma^2_X$. 

\subsection{Instantaneous energy density}
The evolution of the nonstationary energy density is of fundamental importance for incoherent fields. 
When considering the instantaneous field as pure functions of time (i.e., not as an analytic function or phasor), the following definition for the associated energy density applies \cite[p. 14]{papa1}, \cite{kong1}:
\bea
W_e(t) 
&=& \int^t_{-\infty} w_e(t^\prime) {\rm d}t^\prime  \nonumber\\
&\stackrel{\Delta}{=}& \int^t_{-\infty} \ul{\cal Y}(t^\prime) \cdot \frac{\partial \left [ \epsilon(t^\prime) \ul{\cal Y}(t^\prime) \right ]}{\partial t^\prime} \rmd t^\prime
= \frac{1}{2} \epsilon {\cal Y}^2(t)~~~
\eea
where the latter equality holds for a homogeneous time-invariant medium [$\ul{\cal D}(t) \stackrel{\Delta}{=} \epsilon(t) \star \ul{\cal E}(t) = \epsilon \ul{\cal E}(t)$], as we shall further assume.

The previous analysis for a mean complex field $Y(t)$ holds exactly if $Y(t)$ instead symbolizes the mean intensity of a complex field. If we maintain the original definition, however, then
for anharmonic modulated fields that are being perturbed relatively slowly with respect to a central frequency $\omega$, we can use the Gabor analytic field representation 
$Y(t) = {\cal Y}(t) \exp ( - \rmj \omega t )$ to write
\bea
w_e(t) = \ul{Y}(t) \cdot \frac{\partial \left [ \epsilon \ul{Y}^*(t) \right ]}{\partial t}.
\label{eq:energy_iso_slow}
\eea
Therefore, for dispersionless $\epsilon$, the quantity $w_e(t)$ for analytic fields is proportional to the field intensity $|Y(t)|^2$. 
The general expression of $W_e(t)$ for nonstationary random fields is derived in the Appendix
 and is given by (\ref{eq:final_We})--(\ref{eq:I4}).

For the magnetic energy density $W_h$, similar expressions follow by replacing $\epsilon$ by $\mu$, with ${\cal Y}(t)$ now representing the magnetic analytic field.

\subsection{Structure function}

Random field increments provide a transition between stationary systems and fully-developed nonstationary systems (Bachelier--Einstein--Wiener--L\'{e}vy processes). 
They may also be useful for describing undermoded fields, in which small rapid variations ``ride on top'' of a slowly varying mean value. 
Their second-order properties are characterized in general by the structure function, which describes the covariance of field increments \cite{chri1,yagl1}.
It follows upon substitution of Eq. (\ref{eq:Ynoniter2}) as
\bea
&~& \hspace{-0.8cm} D(m,n)\nonumber\\
&\stackrel{\Delta}{=}&
\left \langle \left [ Y(t_0+m \Delta t) - Y(t_0) \right ] \left [ Y(t_0+n \Delta t) - Y(t_0) \right ]^* \right \rangle\nonumber\\
&=&
| Y(t_0) |^2
\left [ 1 - \exp \left ( -\frac{t_m-t_0}{\tau} \right ) \right ]  
\nonumber\\
&~& \times \left [ 1 - \exp \left ( -\frac{t_n-t_0}{\tau} \right ) \right ] + 
\int^\infty_{-\infty} \left [ {1 + (\omega \tau)^2} \right ]^{-1} \nonumber\\
&~& \times \left [ \exp \left ( \rmj \omega t_{m-1} \right )
      - \exp \left ( - \frac{t_{m-1}}{\tau} \right )
      \right ]\nonumber\\
&~& \times
\left [ \exp \left ( \rmj \omega t_{n-1} \right )
      - \exp \left ( - \frac{t_{n-1}}{\tau} \right )
      \right ]
\rmd F_X(\omega).
\label{eq:structfunc}
\eea
In writing the second term in Eq. (\ref{eq:structfunc}), we made use of the fact that $t_{m-1}\simeq t_m$ when $\Delta t/t_0 \rightarrow 0$, and likewise for $t_{n-1}$.
For $m=n$, Eq. (\ref{eq:structfunc}) reduces to
\bea
D(m)
&=&
| Y(t_0) |^2
\left [ 1 - \exp \left ( -\frac{t_m-t_0}{\tau} \right ) \right ]^2
\nonumber\\
&~& + 
\int^\infty_{-\infty} \left [ {1 + (\omega \tau)^2} \right ]^{-1} \nonumber\\
&~& \times \left [ 1-2 \exp \left ( - \frac{t_{m-1}}{\tau} \right ) \cos \left ( \omega t_{m-1} \right ) \right . \nonumber\\
&~& \left.
      + \exp \left ( - \frac{2 t_{m-1}}{\tau} \right ) \right ]
\rmd F_X(\omega),
\label{eq:structfunc_mequaln}
\eea
which is formally equivalent to (\ref{eq:nineteen}). For example, for an Ornstein--Uhlenbeck process, (\ref{eq:structfunc_mequaln}) becomes
\bea
D(m) = \frac{2\sigma^2_0{\cal T}}{\tau+{\cal T}} \left [ 1 - \exp \left ( - \frac{2 t_{m-1}}{\tau} \right ) \right ].
\eea

\subsection{Example: EMC immunity testing using nonstationary mode-stirred vs. quasi-stationary mode-tuned reverberation fields\label{sec:applic}}
The above analysis can be applied to determine the response of a linear DUT that is characterized by an impulse response function $h(t)$ or a frequency characteristic $H(\omega)$ and which has been placed inside a mode-stirred reverberation chamber.
At any location inside the cavity, the local field consists of multipath reflections, i.e., plane waves arriving from isotropically distributed directions with randomly polarized direction of polarization and uniformly distributed phases. 
Within a time interval $\delta t$ during a mode stirring process, each plane wave undergoes a transition of its directions of arrival, direction of polarization and absolute phase. The transition of each parameter can be represented by a trajectory in state space. The actual (i.e., physical) rate of fluctuation during this transition is a function of the rate of change of cavity perturbation and degree of chaoticity of the cavity. The perceived rate is, in addition, a function of the characteristic time (response time) of the DUT to changes of the excitation field.

In general, the specific form of $H(\omega)$ for an DUT is often unknown. Here, as in \cite{arnaTEMCv47n4}, we simply consider a class of DUTs and investigate to what extent a mode-stirred (i.e., continuously varying) vs. mode-tuned (i.e., discrete, stepwise changing) excitation field has a difference in effect on the response of the DUT. Like the excitation, this response is a random function, hence comparison between both cases requires determination of one or more statistical metrics, such as the mean, $\eta$\%-confidence intervals, statistics of the maximum-to-mean ratio, upward threshold crossing frequency, excursion length, etc.

We shall consider the case of an Ornstein--Uhlenbeck process, as a canonical case, for which explicit results can be calculated and from which results for white noise $X(t)$ will follow as a special case. 
Since the (real) power spectrum is
\bea
\rmd F_X(\omega) = f_X(\omega) \rmd \omega = \frac{F_{X,0}}{1+\omega^2 {\cal T}^2} \rmd \omega,
\eea
where $F_{X,0}=Z_{X,0}Z^*_{X,0}$, the (complex) amplitude spectrum is
\bea
\rmd Z_X(\omega) = z_X(\omega) \rmd \omega = \frac{Z_{X,0}}{1+\rmj \omega {\cal T}} \rmd \omega.
\eea
Hence,
\bea
Y(t) 
&=&
Y(t_0) \exp \left ( - \frac{t-t_0}{\tau} \right ) 
+
\int^{+\infty}_{-\infty} \exp \left ( \rm j \omega t \right ) \nonumber\\
&~& \times
\frac{1-\exp \left [ - \left ( 1 + \rmj \omega \tau \right ) \frac{t-t_0}{\tau} \right ]}{(1 + \rmj \omega \tau) (1 + \rmj \omega {\cal T}) } Z_{X,0} \rmd \omega\label{eq:evol_OU}.
\eea
Then the ``output'' field $W(t)$ of the DUT due to a mode-stirred excitation field, i.e., as measured or perceived at one of its test ports, a critical internal component, etc., is then \cite[Sec. 6]{prie1}
\bea
&~& \hspace{-0.8cm}
W(t)  \nonumber\\
&=& \int^{+\infty}_{-\infty} h(u) Y(t-u) \exp \left [ - \rmj \omega_0(t-u) \right ] \rmd u\\
&=& \int^{+\infty}_{-\infty} H_{\omega+\omega_0}(\omega; t) A(\omega+\omega_0;t) 
\exp ( \rmj \omega t) \rmd Z_X(\omega+\omega_0)
\\
&=& \int^{+\infty}_{-\infty} H_{\omega+\omega_0}(\omega; t) 
\exp ( \rmj \omega t)\nonumber\\
&~&\times 
\left [
Y(t_0) \exp \left ( - \frac{t-t_0}{\tau} \right ) \delta(\omega+\omega_0)
\right. \nonumber\\
&~&\left.~~~ +
\frac{1-\exp \left \{ - \left [ 1 + \rmj (\omega+\omega_0) \tau \right ] \frac{t-t_0}{\tau} \right \} }{\left [ 1 + \rmj (\omega+\omega_0) \tau \right ] \left [ 1 + \rmj ( \omega +\omega_0 ) {\cal T} \right ] } Z_{X,0} 
\right ] \rmd \omega
\eea
where $\omega_0$ is any constant frequency and
\bea
H_\lambda(\omega;t) = \int^{+\infty}_{-\infty} h(u) \frac{A(\lambda;t-u)}{A(\lambda;t)} \exp ( - \rmj \omega u ) \rmd u
\eea
at $\lambda = \omega+\omega_0$.

\section{Second-order processes}
The response of an overmoded system can only be characterized approximately and in the mean, as a first-order system. In reality, the large number of participating modes (i.e., those within the instantaneous resonance band of the cavity) and their strong intermodal coupling causes the response to be more irregular, requiring a higher-order description.
To investigate the effect of increasing the order, we here analyze a second-order system. Because of the resonant nature of the modes, this is expected to provide more accurate results. Any higher-order process then follows from cascading of first- and second-order systems.

For the second-order system characterized by
\bea
\frac{\rmd^2 Y}{\rmd t^2} + 2 \zeta \frac{\rmd Y}{\rmd t} + \omega^2_n Y = K_0 \omega^2_0,
\label{eq:ODE_secondorder}
\eea
the step response is the inverse Laplace transform of 
$
Y(t) = {K_0 \omega^2_n}/[{s\left (s^2+2 \zeta s + \omega^2_0\right )}]
$.
The residues of the simple poles $s_0=0$ and $s_{1,2} = -\zeta \pm \rmj \sqrt{ \omega^2_0 - \zeta^2} $ lead to the solution
\bea
Y(t) &=& K_0 
\left \{ 1 - \frac{\omega_0 \exp \left ( - \zeta t \right )}
                  {\sqrt{\omega^2_0 - \zeta^2}} \right.\nonumber\\
&~& \left. \times
\cos \left [ \sqrt{\omega^2_0 - \zeta^2} \thinspace t - \sin^{-1} \left ( \frac{\zeta}{\omega_0} \right ) \right ] 
\right \}
\label{eq:stepresponse_secondorder}
\eea
for $0 <\zeta < \omega_0$ [For overdamped regime ($\zeta > \omega_0$), the step response involves hyperbolic functions.]

In principle, the previous analysis of the first-order model can be repeated. To this end, Eq. (\ref{eq:Ynoniter}) could be written as
\bea
&~& \hspace{-0.8cm} 
Y(t_m \leq t \leq t_{m+1}) 
\nonumber\\
&=&
Y(t_0) \phi^m(\Delta t) \phi(t-t_m) + \phi(t-t_m) 
\nonumber\\
&~&
\times
\left [
1 - \phi(\Delta t)
\right ]
\sum^{m-1}_{i=0} \phi^{m-1-i}(\Delta t) X(t_i)
\nonumber\\
&~&
+
\left [ 1 - \phi(t - t_{m})
\right ] X(t_m)
\label{eq:Ynoniter_bis}
\eea
now with
\bea
\phi(t) = \frac{\omega_0 \exp \left ( - \zeta t\right )}{\sqrt{\omega^2_0-\zeta^2}} \cos \left [ \sqrt{\omega^2_0-\zeta^2} \thinspace t - \sin^{-1}\left ( \frac{\zeta}{\omega_0} \right ) \right ]
\label{eq:phi_order2}
\eea
in view of $\omega_n \simeq \omega$, $\tau_n \simeq \tau$.
To obtain the spectral expansion, it is advantageous to consider the Gabor analytic signal representation $Y(t) = Y^\prime(t) +\rmj Y^{\prime\prime}(t)$. The step response 
is then
\bea
Y(t) &=& K_0 
\left \{ 1 - \frac{\omega_0 \exp \left ( - \zeta t \right )}
                  {\sqrt{\omega^2_0 - \zeta^2}} 
\right. \nonumber\\ &~& \left. \times
\exp \left [ \rmj \left ( \sqrt{\omega^2_0 - \zeta^2} \thinspace t - \sin^{-1} \left ( \frac{\zeta}{\omega_0} \right )
\right ) \right ]
\right \}.
~~~~
\eea
Because of the presence of the factor $1/\sqrt{\omega^2_0-\zeta^2}$ in Eq. (\ref{eq:stepresponse_secondorder}), this approach does not lend itself to express $Y(t)$ easily as an oscillatory or related process, in the spirit of Eqs. (\ref{eq:Yintegral_zerostart})--(\ref{eq:defphi}) for the first-order system.
Therefore, we write Eq. (\ref{eq:ODE_secondorder}) instead as a system of two coupled first-order equations:
\bea
\left \{
\begin{array}{c}
\frac{\rmd Y(t)}{\rmd t} + \frac{1}{\tau_1} Y(t) =  \frac{1}{\tau_1} X(t) \\ 
\\
\frac{\rmd Z(t)}{\rmd t} + \frac{1}{\tau_2} Z(t) =  \frac{1}{\tau_2} Y(t) , 
\end{array}
\right .
\label{eq:LangevinChamber}
\eea
thereby assuming that
\bea
\tau_1 \ll \tau_2 \ll 1
\label{ineq:chamber}
\eea
where the response is now given by $Z(\omega)$, with $Y(\omega)$ denoting an auxiliary intermediate process.
For an overdamped second-order system ($\zeta > \omega_0$), such a decomposition is always possible.
The process $Y(t)$ has a physical meaning: for a first-order system responding to a nonstationary field which, in turn, is governed by a first-order process, the excitation of this system is given by $Y(t)$.
The spectral representation follows by replacing $X(s)$ in Eq. (\ref{eq:six}), as given by Eq. (\ref{eq:seven}) and $Y(s)$ given by Eq. (\ref{eq:evol1}):
\bea
&~& \hspace{-0.6cm} Z(t) \nonumber\\
&=& 
Z(t_0) \exp \left ( - \frac{t-t_0}{\tau_2} \right )
+
\frac{Y(t_0)}{2\zeta \tau_2} \exp \left [
- \left ( \frac{t}{\tau_2} - \frac{t_0}{\tau_1} \right ) \right ]\nonumber\\
&~& \times
\left [ \exp \left ( -2 \zeta t_0 \right ) - \exp \left ( - 2 \zeta t\right ) \right ]\nonumber\\
&~&
+ \int^{+\infty}_{-\infty}
\left \{
\frac{\exp \left ( \rmj \omega t \right ) - \exp \left ( - \frac{t-t_0}{\tau_2} \right ) \exp \left ( \rmj \omega t_0 \right )}{\left ( 1 + \rmj \omega \tau_1 \right ) \left ( 1+ \rmj \omega \tau_2 \right )} \right . \nonumber\\
&~&\left.
+
\frac{\exp \left ( \rmj \omega t_0 \right )}{1+\rmj \omega \tau_1} \frac{\tau_1}{\tau_2} \left [ \exp \left ( - 2 \zeta t + \frac{t_0}{\tau_1} \right )- \exp \left ( - \frac{t}{\tau_2} \right ) \right ] \right \}\nonumber\\
&~& \times
{\rm d} Z_X(\omega).
\eea
For the case where $t_0=0$, $Y(t_0)=Z(t_0)=0$, this expression reduces to
\bea
Z(t)
&=& 
\int^{+\infty}_{-\infty}
\left \{
\frac{\exp \left ( \rmj \omega t \right ) - \exp \left ( - \frac{t}{\tau_2} \right ) }{\left ( 1 + \rmj \omega \tau_1 \right ) \left ( 1+ \rmj \omega \tau_2 \right )} +
\frac{1}{1+\rmj \omega \tau_1} \frac{\tau_1}{\tau_2} \right . \nonumber\\
&~&\left.
\times \left [ \exp \left ( - 2 \zeta t \right )- \exp \left ( - \frac{t}{\tau_2} \right ) \right ] \right \}
{\rm d} Z_X(\omega).
\eea
For an Ornstein--Uhlenbeck process, contour integration yields
\bea
Z(t) &=& -\rmj \frac{2\sigma^2_0{\cal T}^2}{\tau+{\cal T}} \left \{ \frac{{\cal T}}{{\cal T}+ \tau_2} \left [ \exp \left ( - \frac{t}{{\cal T}} \right ) + \exp \left ( - \frac{t}{\tau_2}\right ) \right ] \right. \nonumber\\ 
&~& \left. - \frac{\tau_1}{\tau_2} \left [ \exp \left ( - 2 \zeta t \right ) - \exp \left ( - \frac{t}{\tau_2} \right ) \right ] \right \}.
\eea
Note that, even though the original $\omega_0$ and $\zeta$ have been replaced by $\tau_1$ and $\tau_2$, exhibiting a one-to-one relationship, the process of replacing the second-order SDE by two coupled first-order SDEs requires for one time constant to be much smaller than the other one, in order to maintain the validity of the Langevin--It\^{o} formulation. 

\section{Conclusions}
Angular spectral plane-wave expansions provide a framework for the stochastic characterization of dynamic random EM fields that result from moving boundaries, spatial scanning, and/or changing excitation frequency with consequent changes in the pattern of participating eigenmodes.
It was shown that the spectral expansion can be expressed as an oscillatory process Eq. (\ref{eq:Yintegral_zerostart}) with kernel Eq. (\ref{eq:defphi}) (or, more generally, as Eqs. (\ref{eq:evol1}), (\ref{eq:Yintegral_generalsol}), (\ref{eq:theta})).
The fact that the expansion can still be represented by an oscillatory process (which is usually associated with narrowband, i.e., slowly modulating (quasi-stationary) processes, but now with no such limitation on the rate of fluctuations) can be attributed to the linearity of the system. The results for highly overmoded resonances apply in the mean (incoherent superposition), i.e., for the collection of cavity modes participating in forming the instantaneous field at any one time. 

A useful extension would be the "microscopic" plane-wave spectral expansion for a single second-order resonant system (cavity mode), without any limiting condition on the separation of the two time constants involved in decomposing into two first-order processes. Such a complete second-order characterization could form the basis for a coherent superposition of excited cavity modes.


\clearpage

\appendix
\section*{Appendix: Instantaneous energy density}
Here, we derive the general expression for the instantaneous energy of a nonstationary field.
Substitution of the general expression Eq. (\ref{eq:evol1}) for $Y(t)$ into Eq. (\ref{eq:energy_iso_slow}) yields
\bea
w_e (t^\prime)
&=& 
- \frac{\epsilon |Y(t_0)|^2}{\tau} \exp \left [ - \frac{2(t^\prime-t_0)}{\tau} \right ] 
\nonumber\\
&~&
- 
\frac{\epsilon Y^*(t_0)}{\tau} \exp \left ( - \frac{t^\prime-t_0}{\tau} \right ) 
\int^{+\infty}_{-\infty} \exp \left ( \rmj \omega t^\prime \right ) \frac{ 1 - \exp \left [ - \left ( 1 + \rmj \omega \tau \right ) \frac{t^\prime-t_0}{\tau} \right ]}{1+\rmj \omega \tau} {\rm d} Z_X(\omega)\nonumber\\
&~&
+
\frac{\epsilon Y(t_0)}{\tau} \exp \left ( - \frac{t^\prime-t_0}{\tau} \right ) 
\int^{+\infty}_{-\infty} 
\exp \left ( -\rmj \omega t^\prime \right ) \frac{ \left \{ - \rmj \omega {\tau} + \exp \left [ - \left ( 1 - \rmj \omega \tau \right ) \frac{t^\prime-t_0}{\tau} \right ]\right \}}{1-\rmj \omega \tau} {\rm d} Z^*_X(\omega)\nonumber\\
&~& 
+ \frac{\epsilon}{\tau}
\int^{+\infty}_{-\infty} \int^{+\infty}_{-\infty} 
\exp \left [ \rmj \left ( \omega_1 - \omega_2 \right ) t^\prime \right ] 
\frac{1-\exp \left [ - \left ( 1 + \rmj \omega_1 \tau \right ) \frac{t^\prime-t_0}{\tau} \right ]}{1+\rmj \omega_1 \tau} \nonumber\\
&~&~~~~~~~~~~~~~~\times
\frac{\left \{ -\rmj \omega_2 \tau + \exp \left [ - \left ( 1 - \rmj \omega_2 \tau \right ) \frac{t^\prime-t_0}{\tau} \right ] \right \} }{1-\rmj \omega_2 \tau}
\rmd Z_X(\omega_1) \rmd Z^*_X(\omega_2). 
\label{eq:energy_iso_slow_explicit}
\eea
Integration of $w_e(t^\prime)$ yields the instantaneous $W_e(t)$ as
\bea
W_e(t) &=& \int^t_{t_0} w_e(t^\prime) {\rm d}t^\prime 
= W_{e_1}(t) + W_{e_2}(t) + W_{e_3}(t) + W_{e_4}(t)
~~~~~~~~~~~\label{eq:final_We}
\eea
with
\bea
W_{e_1}(t) &=& -\frac{\epsilon |Y(t_0)|^2}{2} \left \{ 1 - \exp \left [ - \frac{2(t-t_0)}{\tau} \right ] \right \}\label{eq:final_We1}\\
W_{e_2}(t) &=& \epsilon Y^*(t_0) \int^{+\infty}_{-\infty} \frac{\exp \left ( \rmj \omega t_0 \right )}{1+\omega^2 \tau^2} \left \{ \exp \left [ - \left ( 1-\rmj \omega \tau \right ) \frac{t-t_0}{\tau} \right ] - \exp \left [ - \frac{2(t-t_0)}{\tau} \right ] \right \}{\rm d}Z_X(\omega)~~~\\
W_{e_3}(t) &=& \frac{\epsilon Y(t_0)}{2} \left \{ 1- \exp \left [ - \frac{2(t-t_0)}{\tau} \right ] \right \} \int^{+\infty}_{-\infty} \frac{\exp \left ( \rmj \omega t_0 \right )}{1 - \rmj \omega \tau} {\rm d}Z^*_X(\omega)\nonumber\\
&~& - \epsilon Y(t_0) \int^{+\infty}_{-\infty} \frac{\rmj \omega \tau \exp \left ( -\rmj \omega t_0 \right ) }{1+ \omega^2 \tau^2} \left \{ 1 - \exp \left [ - \left ( 1 + \rmj \omega \tau \right ) \frac{t-t_0}{\tau} \right ] \right \} {\rm d}Z^*_X(\omega)\\
W_{e_4}(t) &=& \epsilon \int^{+\infty}_{-\infty} \int^{+\infty}_{-\infty} \frac{I_1(t;\omega_1,\omega_2)+I_2(t;\omega_1,\omega_2)+I_3(t;\omega_1,\omega_2)+I_4(t;\omega_1,\omega_2)}{(1+\rmj \omega_1 \tau)(1 -\rmj \omega_2 \tau)} {\rm d}Z_X(\omega_1) {\rm d}Z^*_X(\omega_2) ~~~~~~~\label{eq:final_We4}
\eea
and
\bea
I_1(t;\omega_1,\omega_2) &=& \frac{\omega_2}{\omega_1-\omega_2} \exp \left [ \rmj (\omega_1 - \omega_2) t_0 \right ] \left \{ 1 - \exp \left [ \rmj (\omega_1-\omega_2) (t-t_0)\right ]\right \} ~~~~~~\label{eq:I1} \\
I_2(t;\omega_1,\omega_2) &=& \frac{1}{1-\rmj \omega_1 \tau} \exp \left (- \frac{2 t_0}{\tau} \right ) \exp \left [ \rmj (\omega_1 - \omega_2) t_0 \right ] \left \{ 1 - \exp \left [ - \left ( 1 - \rmj \omega_1\tau \right ) \frac{t-t_0}{\tau}\right ]\right \} ~~~~~~\label{eq:I2} \\
I_3(t;\omega_1,\omega_2) &=& \frac{\rmj \omega_2 \tau }{1+\rmj \omega_2 \tau} \exp \left [ \rmj (\omega_1 - \omega_2) t_0 \right ] \left \{ 1 - \exp \left [ - \left ( 1 + \rmj \omega_2\tau \right ) \frac{t-t_0}{\tau}\right ]\right \} ~~~~~~\label{eq:I3} \\
I_4(t;\omega_1,\omega_2) &=& - \frac{1}{2} \exp \left ( - \frac{2 t_0}{\tau} \right ) \exp \left [ \rmj (\omega_1 - \omega_2) t_0 \right ] \left \{ 1 - \exp \left [ - \frac{2(t-t_0)}{\tau}\right ]\right \}. ~~~~~~\label{eq:I4}
\eea
For $\omega_1=\omega_2\stackrel{\Delta}{=}\omega$,
\bea
I_1(t;\omega) &=& -\rmj \omega (t-t_0)\label{eq:I1_0} \\
I_2(t;\omega) &=& \frac{1}{1-\rmj \omega \tau} \exp \left (- \frac{2 t_0}{\tau} \right ) \left \{ 1 - \exp \left [ - \left ( 1 - \rmj \omega \tau \right ) \frac{t-t_0}{\tau}\right ]\right \} ~~~~~~\label{eq:I2_0} \\
I_3(t;\omega) &=& \frac{\rmj \omega \tau }{1+\rmj \omega \tau} \left \{ 1 - \exp \left [ - \left ( 1 + \rmj \omega\tau \right ) \frac{t-t_0}{\tau}\right ]\right \} ~~~~~~\label{eq:I3_0} \\
I_4(t;\omega) &=& - \frac{1}{2} \exp \left ( - \frac{2 t_0}{\tau} \right ) \left \{ 1 - \exp \left [ - \frac{2(t-t_0)}{\tau}\right ]\right \}. ~~~~~~\label{eq:I4_0}
\eea
Considering that $\langle {\rm d}Z(\omega) \rangle = 0$ and $\langle {\rm d}Z(\omega_1) {\rm d}Z^*(\omega_2) \rangle = \delta (\omega_1-\omega_2) {\rm d}F_X(\omega)$, we have that
\bea
\langle W_{e_2} \rangle &=& \langle W_{e_3} \rangle = 0\\
\langle W_{e_4} \rangle &=& \epsilon \int^{+\infty}_{-\infty} \frac{I_1(t;\omega)+I_2(t;\omega)+I_3(t;\omega)+I_4(t;\omega)}{1+\omega^2 \tau^2} {\rm d}F_X(\omega).~~~~~~~~
\eea

\end{document}